\documentclass[twocolumn,twoside,preprintnumbers,amsmath,amssymb,showkeys]{revtex4}
\usepackage{epsfig}
\usepackage{graphicx}

\usepackage{fancyhdr}

\usepackage{pslatex}

\begin{document}

\title{Gluon dominance model and cluster production}

\author{Elena Kokoulina$^{1,2}$, Andrey Kutov$^{2,3}$ and Vladimir Nikitin$^2$}

\affiliation{$^1$ GSTU, October Avenue, 48, Gomel,
       246746, Belarus\\
         $^2$ JINR, Dubna, Moscow region, 141980, Russia \\
         $^3$ DM UrD RAS, Chernova, 3a, Syktyvkar, 167982, Russia }


\begin{abstract}
Gluon dominance model (GDM) studies multiparticle production in
lepton and hadron processes. It is based on the QCD and
phenomenological scheme of hadronization. The model describes well
multiplicity distributions and their moments. It has revealed an
active role of gluons in multiparticle production, it also has
confirmed the fragmentation mechanism of hadronization in $e^+e^-$
annihilation and its change to recombination mechanism in hadron
and nucleus interactions. The GDM explains the shoulder structure
of multiplicity distributions. The agreement with $Au$+$Au$
peripheral collisions data for hadron-pion ratio has been also
obtained with this model. Development of GDM allows one to
research the multiplicity behavior of $p\bar p$ annihilation at
tens of GeV. The mechanism of soft photons production and
estimates of their emission region have been offered. The
experimental data (project "Thermalization", U-70, IHEP) have
confirmed a cluster nature of multiparticle production.


\end{abstract}
\maketitle

\thispagestyle{fancy}

\setcounter{page}{1}

\section{INTRODUCTION}
Heavy ions collisions (HIC) study at high energies reveal strong
evidences of quark-gluon (QG) plasma production \cite{Sal}. The
behavior of bulk variables at lower energies and also a detailed
study of hadron interactions supply with understanding of the
production mechanism of this new state. At present this analysis
is realized at SPS (CERN) \cite{Blu}. The basic problem of HIC is
to describe the systems, consisting of partons or hadrons.
Experiments at RHIC have confirmed this collective behavior
\cite{Bra}. The qualitative properties of the bulk matter may be
represented with the phase diagram. The calculations at QCD and
different models modify it. This diagram is made for long nuclear
medium. It explains transition from the hadron phase to the QG one
in HIC.

The question appears about the manifestation of this transition in
hadron interactions at high energies. We can continue the analogy
proposed in \cite{Tom} for water (boiling, condensing,
ionization). In the case of the hadron interaction the new formed
medium, named quark-gluon plasma (QGP), won't have such a plenty
of constituents. We consider that the evaporation of single
partons from separate hot pots (cluster sources) in the system of
collided hadrons, leads to the secondary particles production.
This conception was taken as the basis of the Gluon Dominance
Model (GDM) \cite{GDM,GDM1,GDM2,GDM3}. Using of this model allows
one to investigate the following problems for the researchers:
multiparticle production (MP), hadronization and phase
transitions. It is interesting to add that the analysis of MP in
the framework of the other picture based on dissipating energy of
participants \cite{EGS} describes the similarity of the bulk
observable as the mean multiplicity in the hadron (nucleus) and
$e^+e^-$ interactions.

\section{$e^+e^-$ annihilation}
The $e^+ e^-$ annihilation is one of the most suitable processes
for initial study of MP. According to the theory of strong
interactions QCD, it is realized through the production of
$\gamma$ or $Z^0$--boson into two pure quarks:
$e^+e^-\rightarrow(Z^0/\gamma)\rightarrow q\bar q$. Perturbative
QCD can describe the fission process of partons (quarks and
gluons) at high energy, because the strong coupling $\alpha_s$ is
small at high energy. This stage can be called as the stage of
cascade. Towards the end of the fission partons have small
virtuality and must change into hadrons, which we observe. At this
stage we could not apply pQCD. Therefore phenomenological models
are used to describe hadronization (transformation of quarks and
gluons into hadrons). Before it was considered that the production
of hadrons from partons is a universal process. Data obtained at
RHIC are an evidence of the mechanism change in the nuclear medium
in comparison with vacuum.

In 90-er we developed a scheme which joints the quark-gluon
cascade and hadronization into a Two Stage Model (TSM)
\cite{GDM,GDM1}. According to it the multiplicity distributions
(MD) in $e^+e^-$ annihilation are
\begin{equation}
\label{7} P_n(s)=\sum\limits_mP^P_m(s)P_n^H(m),
\end{equation}
where $P_m^P(Y)$ is NBD for partons
\begin{equation}
\label{8} P_m^P(s)=\frac{k(k+1) \dots(k+ m-1)}{m!} \cdot \cdot
\left(\frac{\overline m} {\overline m+k}\right)^{m}\left(
\frac{k}{k+\overline m}\right)^{k},
\end{equation}
and $P_n^H(m)$ - binomial distribution (BD) for hadrons produced
from $m$ partons at the stage of hadronization:
\begin{equation}
\label{11} P_n^H={N_p\choose n}\left(\frac{\overline n^h_p}
{N_p}\right)^n\left(1-\frac{\overline n_p^h} {N_p}\right)^{N_p-n},
\end{equation}
where $\overline n^h_p$ and $N_p$ ($p=q,g$) have the meaning of
average multiplicity and a maximum possible number of secondary
hadrons formed from the parton at the stage of hadronization. To
distinguish hadrons produced from the quark and gluon at the
second stage, parameter $\alpha =N_g/N_q$ was introduced. The main
result of the comparison (\ref{7}) \cite{GDM,GDM1} with the
experimental data was in almost constancy of gluon hadronization
parameteres: $N_g\sim 3$ and $\overline n_g^h\sim 1$ (Fig. 1).
From this result we can confirm the universality and fragmentation
character of the gluon hadronization \cite{BMUL} for $e^+e^-$
annihilation. The fact $\alpha <1$ proves that hadronization of a
gluon jet occurs softer than of the quark one.
\begin{figure}[h!]
     \leavevmode
\centering
\includegraphics[width=1.4in, height=1.4in, angle=0]{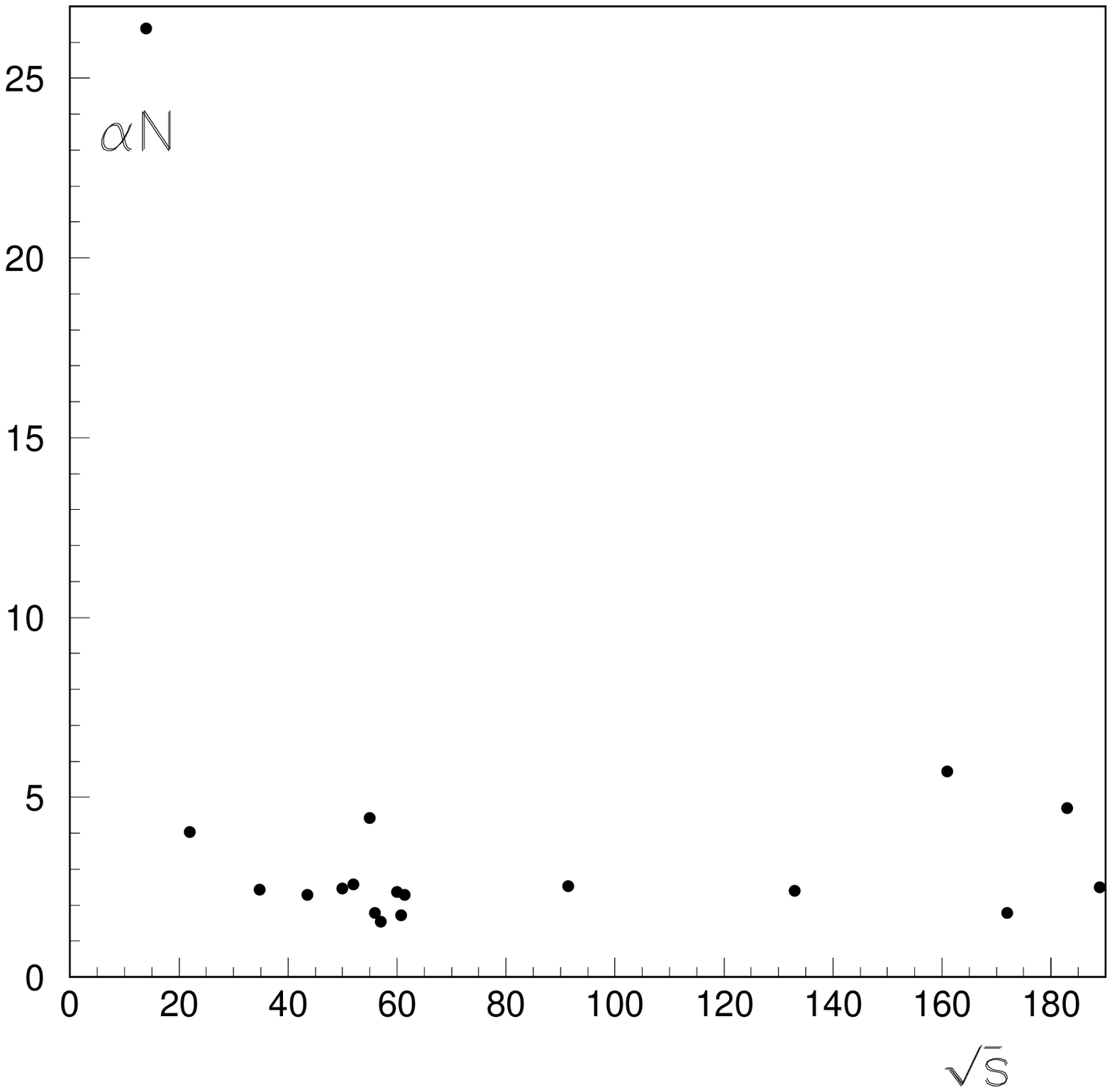}
\includegraphics[width=1.4in, height=1.4in, angle=0]{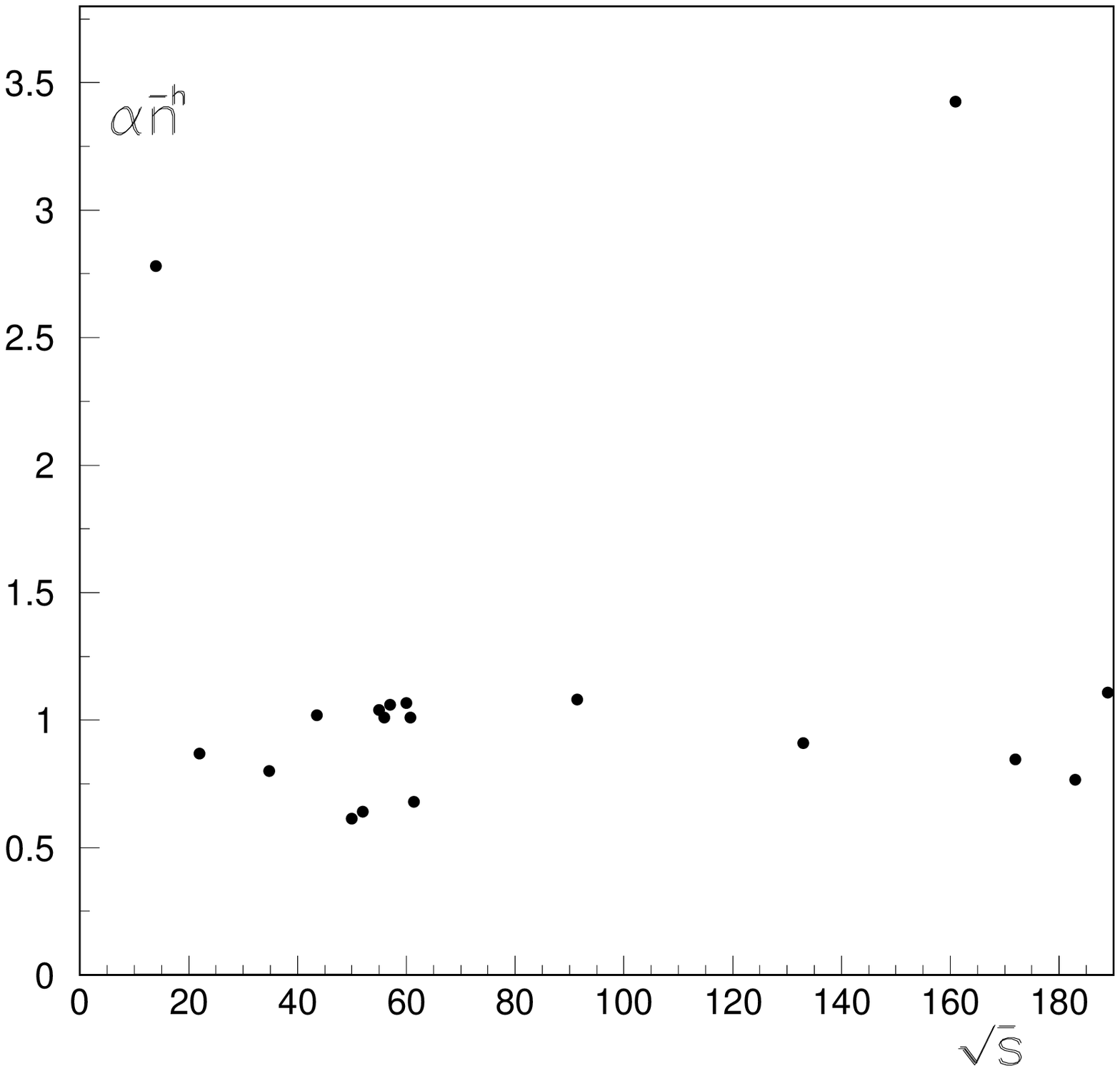}
\caption{The gluon hadronization parameters $N_g$ (left) and
$\overline n_g^h$ (right).}
\end{figure}
We have used MD (\ref{7}) to explain the sign changes as a
function of order for the ratio of factorial cumulative moments
over the factorial ones. In the region lower $Z^0$ this ratio
changes the sign with parity $q$. At higher energies the
oscillation period increases to $4$ and higher. It can be
explained by the influence of the developed cascade and
hadronization \cite{GDM}.

\section{Hadron interactions}
Further development of the TSM scheme and its application to study
proton interactions study has shown an active role of gluons in MP
and confirmed a recombination mechanism of hadronization for them.
That is why this scheme was named as GDM. Our study has shown that
quarks of initial protons stay in leading particles (from 70 to
800 GeV/c and higher). MP is realized by gluons. We call them
"active". In the framework of this model we have madthe following
basic assumptions. concerning the first stage we believe, that
after the inelastic collision of two protons some part of energy
is converted into the thermal (the dissipating energy) and one or
few gluons become free, and they may give a cascade. At the second
stage (hadronization) some of gluons (not all) leave the QG system
-- evaporate and convert to hadrons. Two schemes were offered.

The first scheme takes gluon branch into account. The final MD was
determined as a convolution of three MD: 1) MD of active gluons at
the moment of impact (Poisson), 2) MD of branch gluons (Furry) and
3) MD at the hadronization stage (BD). We have obtained a fraction
of the remained free gluons equal to $0.47 \pm 0.01$ (the same in
\cite{Mul}). The remains of gluons from the QG system can become
sources of soft photons (SP). Moreover, the value of $N_g\approx
40$ was very close to the number of partons in the glob of cold
QGP \cite{LVH}, which explains the production of SP excess
\cite{CHL}.

In the second scheme the gluons leave the QG system (evaporation)
and fragment to hadrons without taking this branch into account.
In this case the final MD are determined by the Poisson (first
stage) and BD (hadronization) convolution
\begin{equation}
\label{38} P_n=\sum\limits_{m}\frac{e^{-\overline m} \overline
m^m}{m!} {mN\choose n-2} \left(\frac{\overline n^h}
{N}\right)^{n-2}\left(1-\frac{\overline n^h} {N}\right)^{mN-n+2}.
\end{equation}
This expression describes the data from 70 to 800 GeV/c $\quad$
(Fig. 2) and gives the gluon hadronization parameters \cite{GDM1}:
$N_g=4.24\pm 0.13$, $\overline n^h_g=1.63\pm 0.12$ at 70 GeV/c.
The gluon hadronization parameter $\overline n^h$ in pp
interactions is comparable with the values obtained in $e^+e^-$
annihilation but has weak growth from 1.63 to 2.66. This behavior
is an evidence of the influence of parton medium on hadronization
in pp interactions, that agrees with the recombination mechanism
when simultaneously few gluons fragment into $q\overline q$ pairs
and form real hadrons at their random walking in the QG system.
Moreover, the maximum number of active gluons $M$ grows from 6 to
10 in this energy region. This value allows one to estimate the
upper limit of multiplicity of the charged hadrons as 26 at 70
GeV/c. MD for neutral mesons and for the total multiplicity were
also obtained in GDM \cite{GDM2}. Using these distributions and
compare them with the data \cite{PIO}, it was shown that the
maximum number of $\pi ^0$ at the charged multiplicity smaller
than the mean one could not be bigger than total number of the
charged particles. The upper limits of neutral and total
multiplicity were defined as well.

The shoulder structure of MD shows up in the region of ISR
energies \cite{ISR} and higher. It can be explained by GDM. As the
active gluons at higher energies may fission, we should take this
into account. The independent evaporation of gluon sources
consists not only of single gluons but groups from two and more
fission gluons as "superposition" with the hadronization
following. This superposition describes well MD at these energies
and gives understanding of soft and (semi)hard components as
clusters with single or few fission gluons, respectively
\cite{GDM3}. The ratio of charged hadron pairs to $\pi ^0$-mesons
was obtained in GDM \cite{GDM3}. It is equal to $\approx 1.6$ and
agrees with the experimental data \cite{PHE}.

We have used the black body emission spectrum at the assumption
that the QG system or excited new formed hadrons are in almost
equilibrium state during a short period of time. The obtained
linear size of SP emission region changes from 4 to 6 fm
\cite{GDM2,Kur}. It is known that hadronization occurs in this
region.

   At modification GDM by the inclusion of intermediate quark
topologies we can describe the experimental differences between
$p\overline p$ and $pp$ inelastic topological cross sections and
second correlation moment behavior at few GeV/c \cite{Rush}. The
tail of high multiplicity in this process originates from "4" or
"6"-topologies.
\begin{figure}[h!]
     \leavevmode
\centering
\includegraphics[width=1.3in, height=1.2in, angle=0]{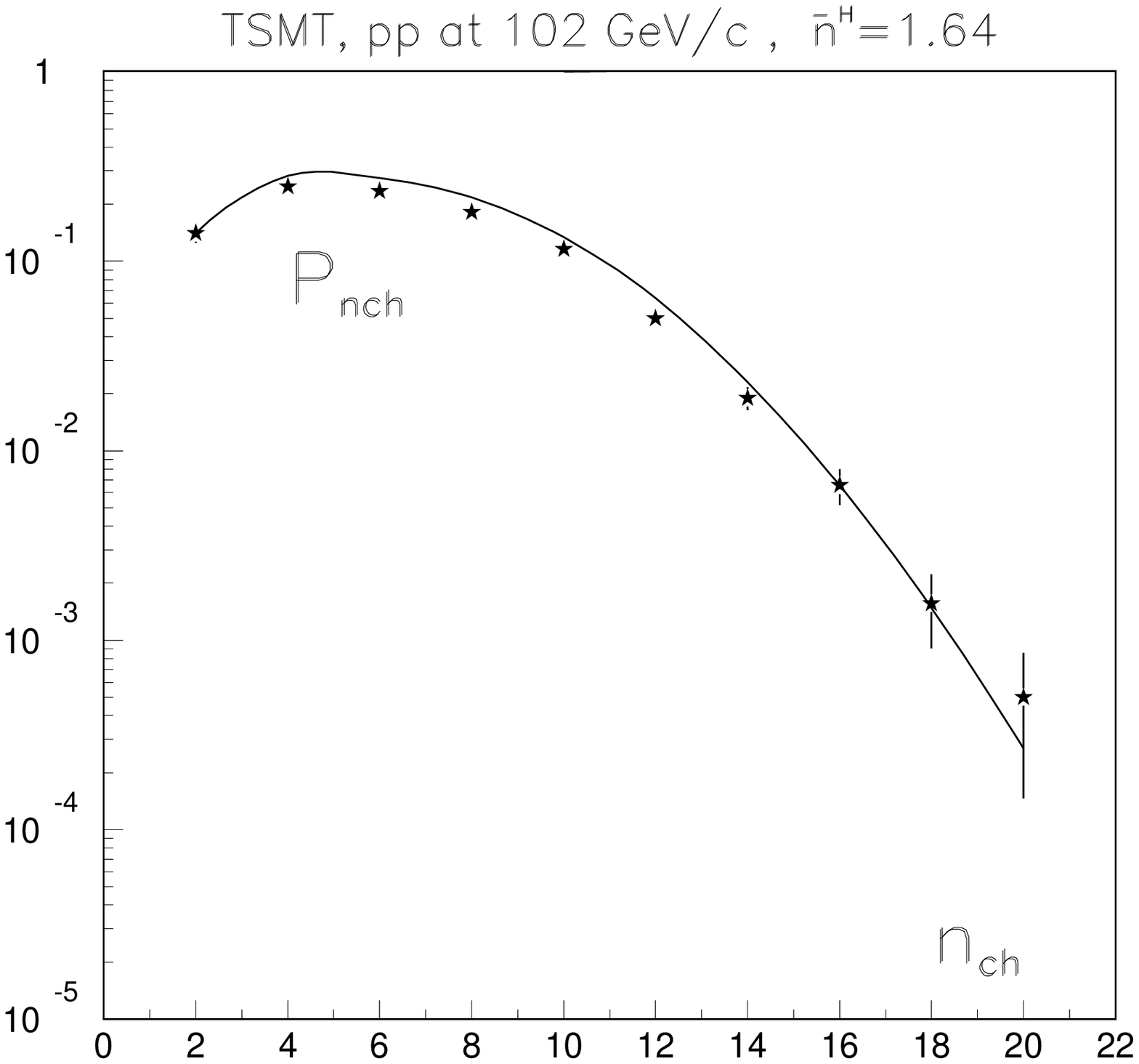}
\includegraphics[width=1.3in, height=1.2in, angle=0]{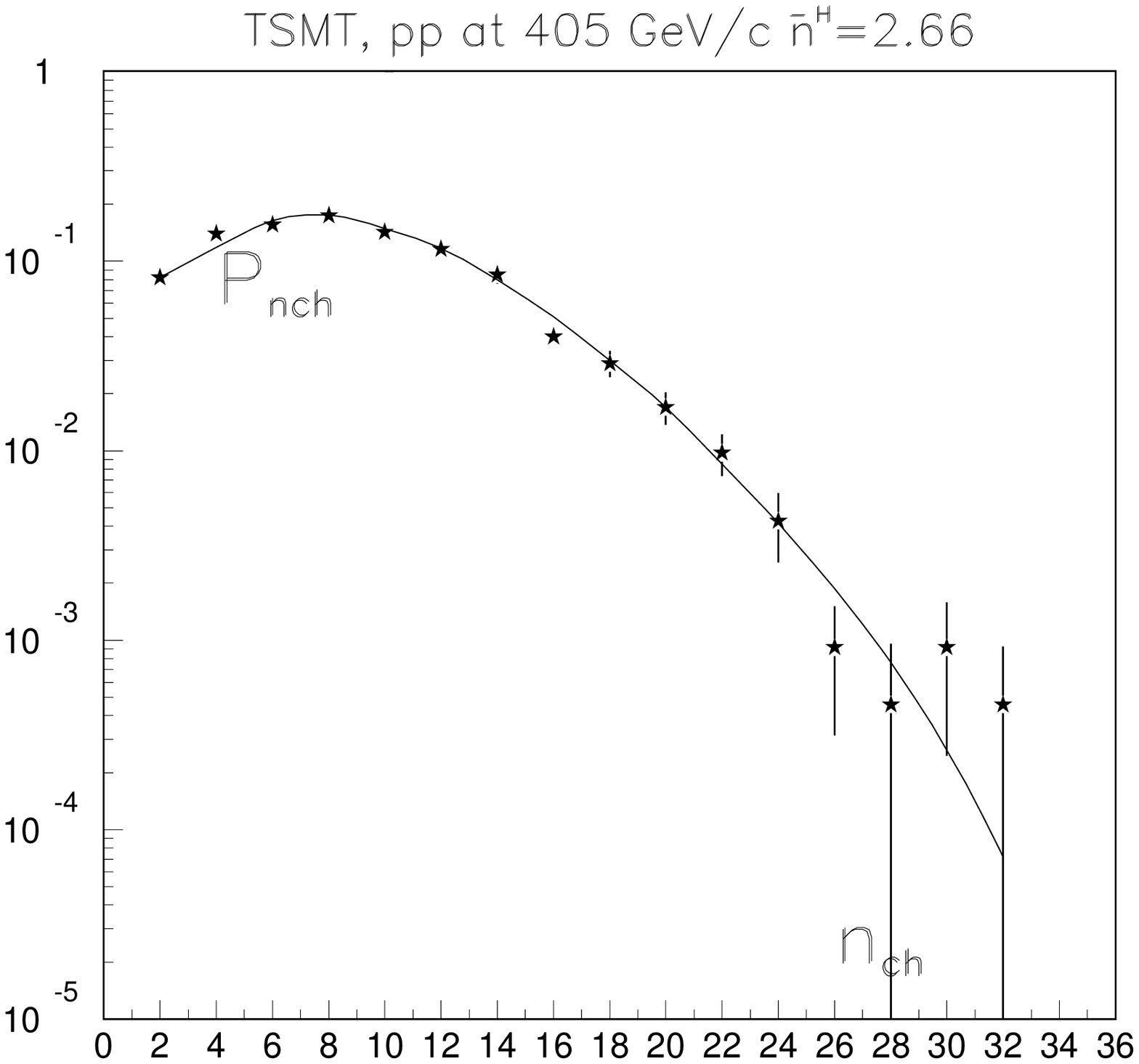}
\caption{GDM MD in pp (solid) (left) at 102 and (right) 405 GeV/c
\cite{Bro}.}
\end{figure}
\section{Project "THERMALIZATION"}

Project "Thermalization" \cite{THERM} is aimed at MP studying the
interaction of the beam energy protons $E_{lab}$=70 GeV of IHEP
(Protvino) with the hydrogen and nucleus targets in the region of
high multiplicity: $n>20$. The experiment is carried out on the
set-up SVD - the Spectrometer with the Vertex Detector supplied
with the trigger system to register rare high multiplicity events.
Our collaboration has designed and manufactured a scintillation
hodoscope or the HM trigger. It should suppress interactions with
track multiplicity below 20. Beyond this it should be thin enough
not to distort the angular and momentum resolutions of the setup
to any kind of the fake signal. The scintillator counter array may
operate at higher counting rate and is more resistant to many
kinds of noise.

    In the region of high multiplicity (HM) $n_{ch}>20$ we expect
\cite{THERM}: formation a high density thermalized hadronic
system, transition to pion condensate or cold QGP, enhanced rate
of SP. We search for new phenomena: Bose-Einstein condensate
(BEC), events with ring topology (Cherenkov gluon radiation),
hadronization, pentaquarks. The available MP models and MC codes
(PYTHIA) are distinguished considerably in the HM region.
\begin{figure}[h!]
\leavevmode \centering
\includegraphics[width=1.5in, height=1.4in, angle=0]{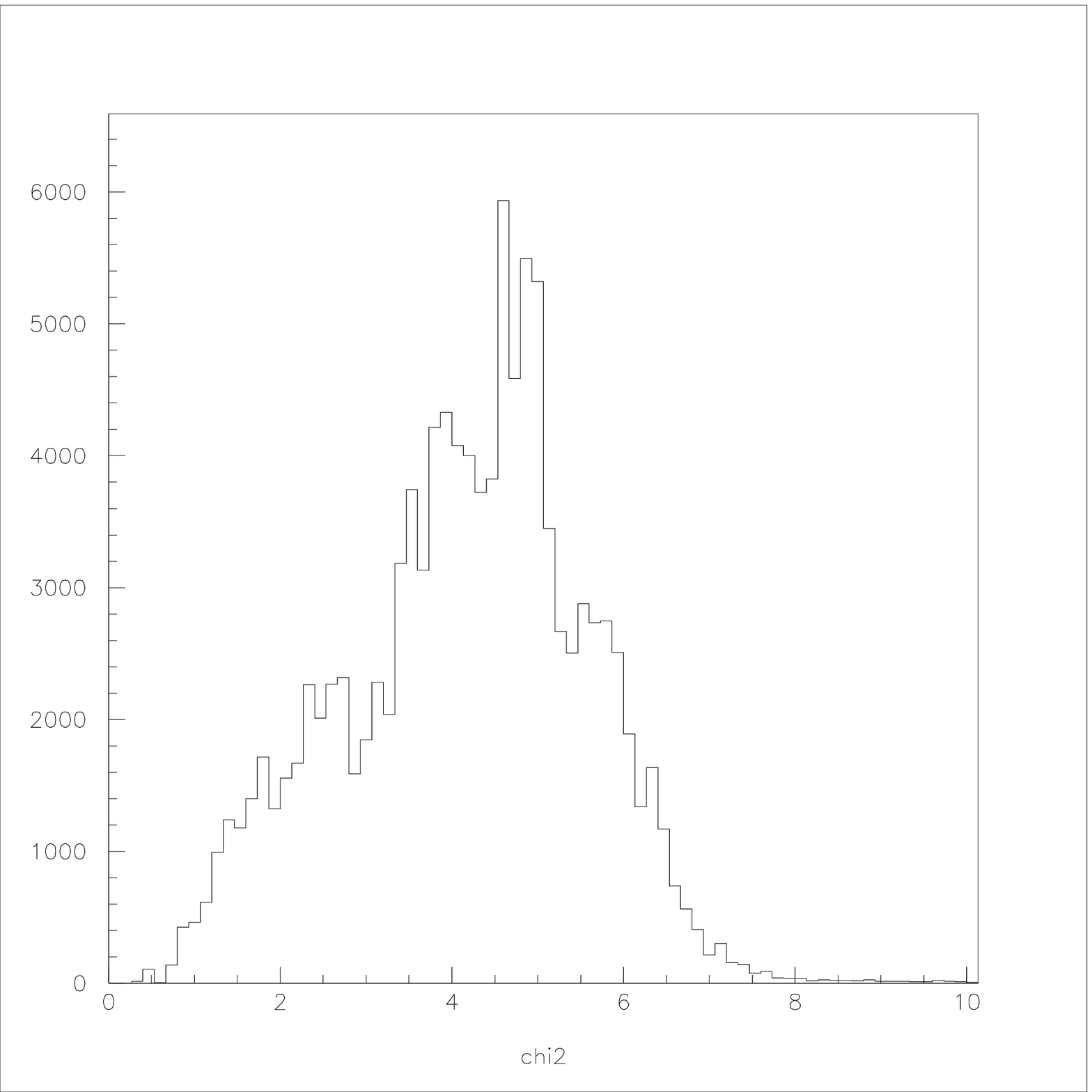}
\includegraphics[width=1.5in, height=1.4in, angle=0]{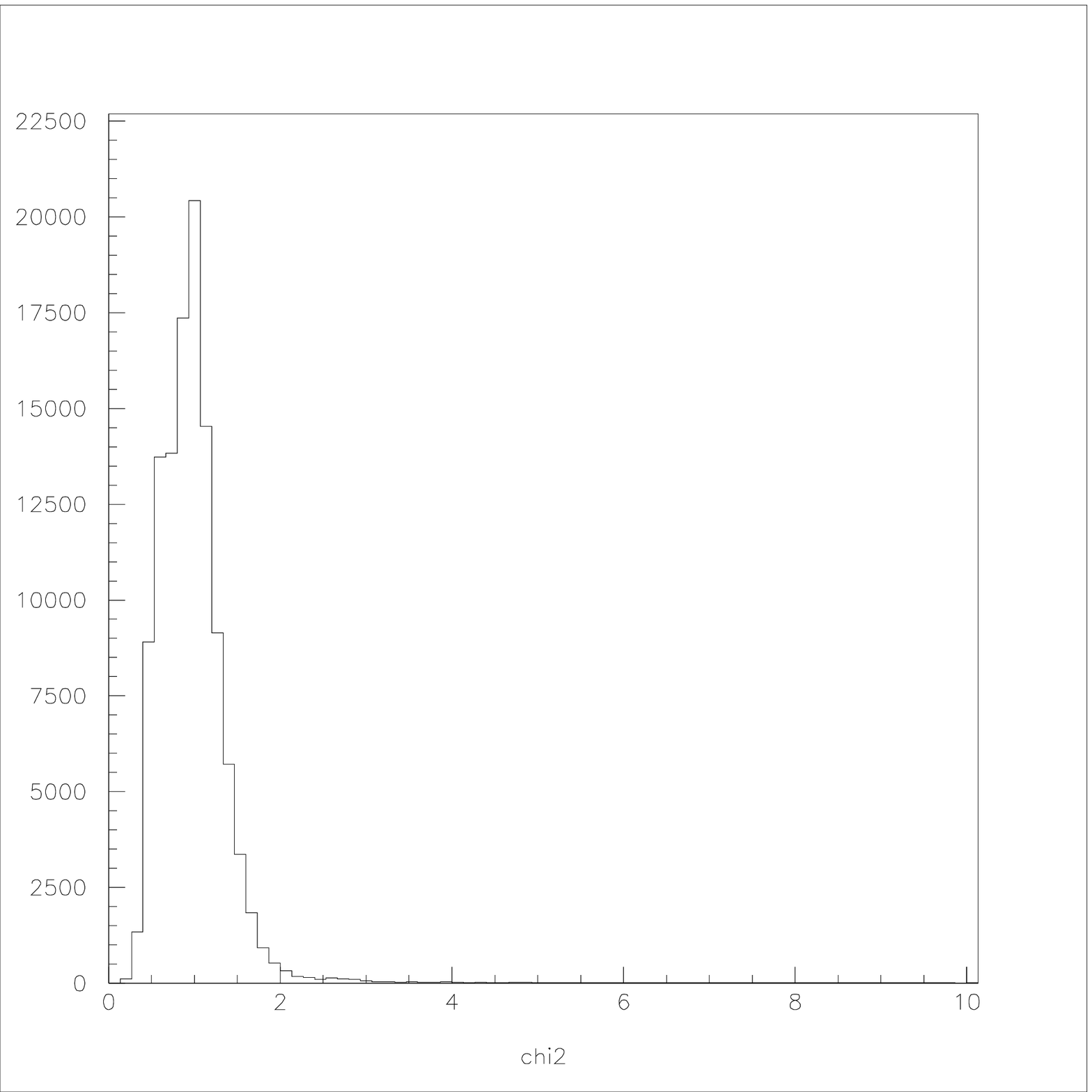}
\caption{$\chi ^2/n_{df}$ for 114769 tracks in magnetic
spectrometer. Left: before alignment, right: after alignment.}
\end{figure}
  In 2005 one technical run was performed at U-70 accelerator with
a hydrogen target and the HM trigger. The alignment task of
detectors had to be solved for the track reconstruction. The
quality check of each detector was estimated through $ \chi ^2$ on
it and residual distributions. For the alignment procedure we have
used a more robust, efficient and high precision method based on
the Linear Least Squares (LLS) with a linear model: $\bf
{u=Aa+r}$, where $\bf {u}$ is a  vector of the measured data, $\bf
{A}$ - matrix, $\bf {a}$ - vector parameters and $\bf {r}$ -
vector residuals. The solution of it $\bf {a=C^{-1} A^T W y}$
requires the inversion of symmetric matrix $\bf {C=A^TWA}$.
V.Blobel \cite{Blob1} designed and applied LLS to the calibration
and alignment tasks. His approach allows one to resolve the
problem with a lot of parameters. Their number may reach thousands
and hundred thousands. V.~Blobel had solved this mathematical
puzzle and put into a general program package $\bf {MILLEPEDE}$
\cite{Blob1} for the efficient solution into practice. We have
developed programme packets based on MILLEPEDE to define
misalignment parameters with good precision (Fig. 3). The
reconstruction of tracks is realized by sequential histogramming
and Hough methods.
\begin{figure}[h!]
\leavevmode \centering
\includegraphics[width=1.5in, height=1.4in, angle=0]{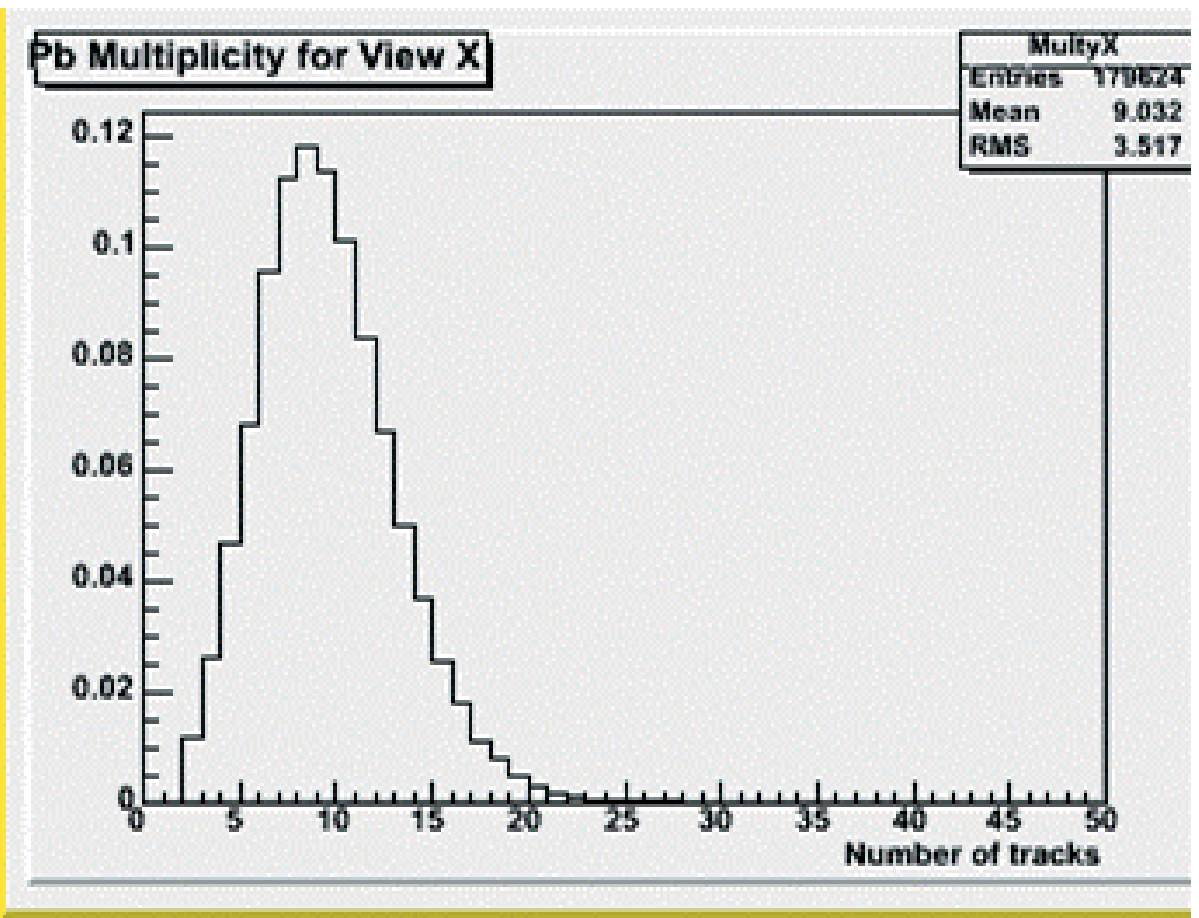}
\includegraphics[width=1.5in, height=1.4in, angle=0]{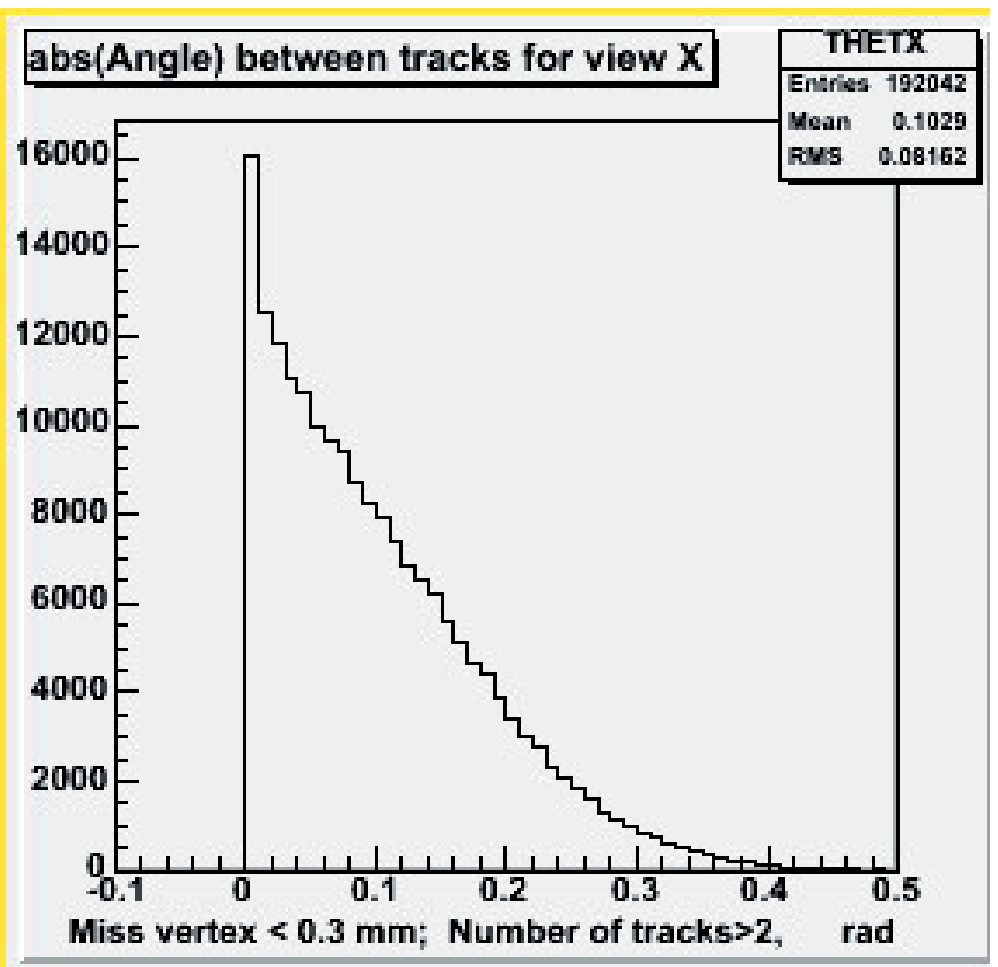}
\caption{Left: MD in $p+Pb$ in x-plane of VD, right: $\Delta
\theta -$ distribution (see text).}
\end{figure}
     The data of the 2002 run for $p+A$ ($A$ = $Si,$ $C$ and $Pb$) were
studied in the HM region, $>20$ charged. MD for these targets were
obtained (Fig. 4, left) by using VD. Interesting phenomenon was
revealed: the indication of grouping of secondary in some certain
direction. Cluster production is seen well -- consists of few
charged (2-4). This is the peak in the differences of the absolute
value of angle $\Delta \theta $ between particles distribution
(Fig. 4, right).

    We are planing to continue this work to make programme packets for
tracking tasks and study new phenomena in the region of high
multiplicity.

\noindent{\bf Acknowledgements}

The implemented study of MP in the framework of {\it project
"Thermalization" is partially supported by RFBR grant
$06-02-81010-Bel\_ a$}.


\end{document}